\documentclass[conference,a4paper]{IEEEtran}
\usepackage{graphicx}
\usepackage{amsmath}
\usepackage{amssymb}
\usepackage{booktabs}
\usepackage[hyphens]{url}
\usepackage[colorlinks=true,allcolors=blue]{hyperref}
\usepackage{algpseudocode}
\usepackage{algorithm}
\usepackage{tabularx}
\usepackage{multirow}
\usepackage{float}
\usepackage{orcidlink}

\title{DDPG-Based Intelligent Handover For STAR-RIS-Assisted 6G Networks}

\author{
    \IEEEauthorblockN{Amr Mansour $^1$\orcidlink{0009-0004-4610-4427}, Hanan Hussein $^1$, Marya Albert $^1$\orcidlink{0009-0004-1552-586X}, Nada Elsharkawy $^2$ , Mohamed Shalma $^1$\orcidlink{0000-0002-7185-1393}}
    \IEEEauthorblockA{$^1$Faculty of Information Engineering and Technology, The German University in Cairo\\
    $^2$Faculty of Engineering, The German International University\\
    Cairo, Egypt\\
    }
}

\begin{document}

\maketitle

\begin{abstract}

Intelligent handover management is a key capability for future 6G networks. This paper investigates enhanced handover strategies in systems assisted by Simultaneously Transmitting and Reflecting Reconfigurable Intelligent Surfaces (STAR-RIS). In addition, it explores the optimal placement of multiple base stations (BSs) to further improve network performance. To eliminate coverage dead zones, a Deep Deterministic Policy Gradient (DDPG) framework is utilized to enforce a Max-Min Fairness objective, directly maximizing the worst-case Signal-to-Noise Ratio (SNR). This optimized topology is evaluated via a dynamic mobility stress test simulating 180 pedestrians executing A3-event handovers under stochastic transient blockages. Results demonstrate that the DDPG deployment outperforms random baselines by a 3.17 dB mean gap, achieving a 24 dB worst-case SNR and a 0.963 Jain’s fairness index. Crucially, this optimal static placement drastically improves dynamic mobility resilience, ensuring a 100\% successful handover rate and reducing post-failure outage time to a negligible 0.01\%.
 
\end{abstract}


\section{Introduction}
Beyond 5G and 6G high-frequency networks face severe propagation challenges in dense urban environments, including high path loss and transient blockages \cite{MacCartney2017}. Reconfigurable Intelligent Surfaces (RISs), specifically Simultaneously Transmitting and Reflecting RISs (STAR-RISs), employing the Energy Splitting (ES) protocol, address these limitations by transmitting and reflecting signals concurrently, achieving full-space coverage and eliminating dead zones \cite{10525765,10525785,Wu2021ChannelSTARRIS}.

Optimal spatial placement of base stations (BSs) and STAR-RIS units in complex, non-convex campus environments is computationally prohibitive for traditional solvers, which often neglect the weakest network links \cite{Shalma2026OptimalBSPlacement}. The Deep Deterministic Policy Gradient (DDPG) algorithm is therefore employed to navigate continuous action spaces and discover exact 2D deployment coordinates \cite{Ullah2025DDPGUAVRIS}. This paper investigates the continuous 2D placement of four BSs and four STAR-RIS units across a non-convex campus, and further evaluates handover resilience under mobility for reliable 6G connectivity.

\subsection{Literature Review}
BS deployment for mmWave systems is strongly dependent on line-of-sight conditions, user distribution, and spatial geometry, requiring careful placement optimization to maintain coverage and data rates \cite{10.1177/1550147720926374}. Recent studies have integrated deep reinforcement learning (DRL) \cite{11232082} for wireless communication. Particularly, the use of DDPG for BS placement and resource optimization rather than fixed mathematical models. Multi-agent DRL and multi-objective formulations have further improved adaptation and jointly optimized coverage, throughput, and localization accuracy in dynamic 5G deployments \cite{10.1109/10597044}.

However, many existing studies rely on simplified propagation assumptions or open-area scenarios, with limited work addressing complex, partially enclosed geometries such as U-shaped environments; robust placement methods addressing demand uncertainty, for instance, typically assume predefined candidate locations and are not tailored to mmWave LOS/blockage dominance \cite{222}. Intelligent DRL-based frameworks are therefore needed for such environments.

These works focus mainly on static BS placement, whereas network performance also depends on maintaining seamless connectivity as users move -- a problem addressed by handover management. Conventional mechanisms follow the 3GPP-standardized A3 event, selecting a candidate cell only once it exceeds the serving cell's signal quality by a hysteresis margin sustained over a Time-to-Trigger (TTT) interval \cite{3GPP36331}. DRL-based approaches have since been explored to improve upon this in high-mobility scenarios where fixed hysteresis/TTT is less effective \cite{Driven}, including proximal-policy-optimization-based adaptive protocols outperforming standard 3GPP handover in data rate and failure rate \cite{Reinforcement}, and online Kalman-filter/SARSA-based TTT-hysteresis adaptation reducing failures under high mobility \cite{Adaptive}.

\subsection{Contribution}
Existing literature lacks a unified framework that jointly optimizes multi-BS/STAR-RIS placement under non-convex spatial constraints and evaluates handover resilience against transient blockages. Addressing these gaps, the contributions of this paper are summarized as follows:
\begin{itemize}
\item A DDPG-driven 2D spatial optimization framework is formulated to determine exact BS and STAR-RIS coordinates, maximizing worst-case user SNR via a strict Max-Min Fairness objective.
\item The STAR-RIS Energy Splitting (ES) protocol is integrated into a dynamic mobility stress test with 180 pedestrians to evaluate the resilience of the standard 3GPP A3-event handover mechanism under transient blockage conditions.
\item The DDPG-driven deployment is shown to outperform random baselines by a 3.17~dB mean gap while achieving a robust 24~dB worst-case SNR; on top of this layout, the handover simulation achieves a 99.5\% successful handover rate, with post-failure outage time reduced to a negligible 0.01\%.
\end{itemize}

\section{System and Channel Model}
\label{sec:System}

\subsection{Model Description}
\begin{figure}[t]
    \centering
    \includegraphics[width=\linewidth]{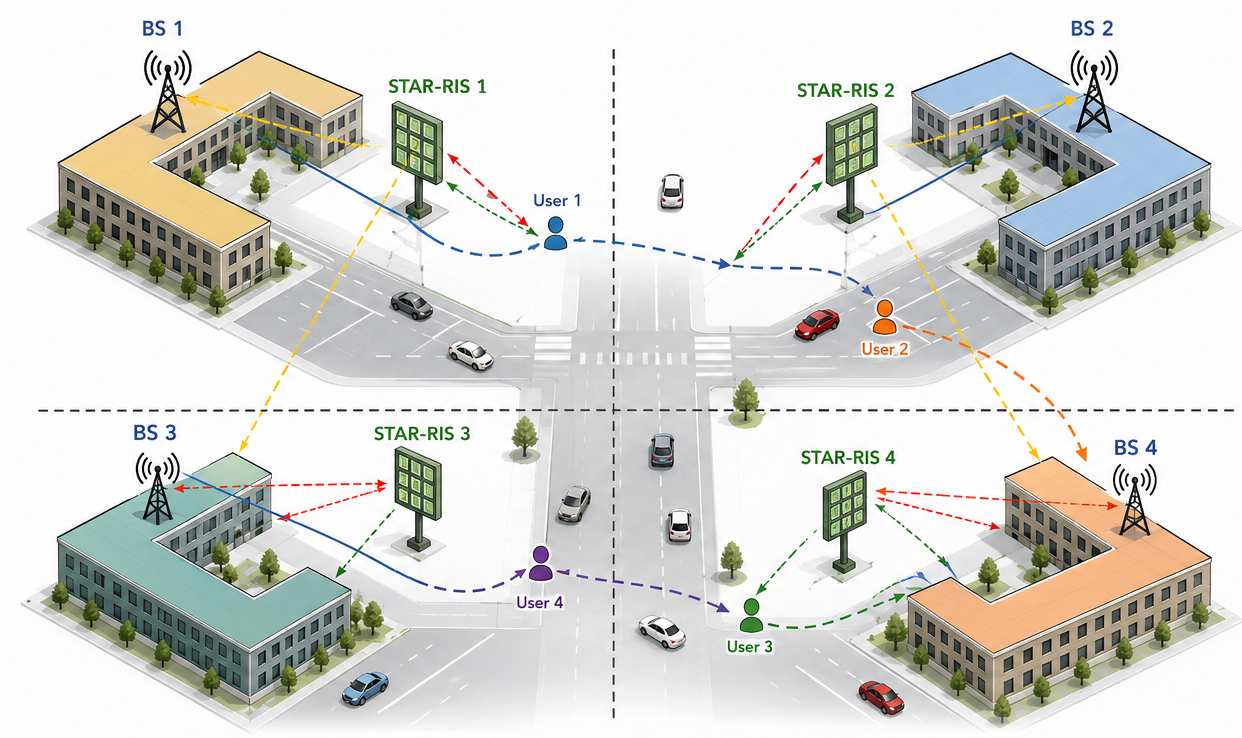}
    \caption{System model configuration}
    \label{fig_sys_mod}
\end{figure}

We consider a downlink Beyond 5G (B5G) mmWave communication scenario. The environment consists of $M=4$ Base Stations (BS) and $M=4$ Simultaneously Transmitting and Reflecting Reconfigurable Intelligent Surfaces (STAR-RIS) serving $K=180$ mobile User Equipments (UEs) as shown in Fig. \ref{fig_sys_mod}.  The BS vicinity is a realistic urban scenario at the German University in Cairo (GUC) campus which is a non-convex topology. The objective is to determine the optimal two-dimensional (2D) continuous coordinates for both the BSs and STAR-RISs nodes to maximize the Max-Min Fairness of the network, subject to physical building constraints and a minimum hardware separation distance.
\subsection{Channel Model}
The Euclidean 3D distance between the base station and user $k$ is given by
\begin{equation}
d_k = \sqrt{(x_{\text{BS}} - x_k)^2 + (y_{\text{BS}} - y_k)^2 + (h_{\text{BS}} - h_k)^2}
\end{equation}
where $(x_{\text{BS}}, y_{\text{BS}})$ and $(x_k, y_k)$ denote the horizontal coordinates of the BS and user $k$, respectively, while $h_{\text{BS}}$ and $h_k$ represent their heights. The mmWave channel between the BS and user $k$ is modeled as a combination of large-scale path loss and small-scale fading:
\begin{equation}
h_k = \sqrt{L(d_k)} \, g_k
\end{equation}
where $g_k \sim \mathcal{CN}(0,1)$ represents Rayleigh fading. The large-scale path loss is modeled as
\begin{equation}
L(d_k) = C_0 d_k^{-\alpha}
\end{equation}
where $C_0$ is the path loss at a reference distance of 1 meter, and $\alpha$ is the path loss exponent. The received signal at user $k$ is expressed as
\begin{equation}
y_k = \sqrt{P_{\text{tx}}} \, h_k x + n_k
\end{equation}
where $P_{\text{tx}}$ is the transmit power, $x$ is the transmitted symbol with unit power, and $n_k \sim \mathcal{CN}(0, N_0)$ denotes additive white Gaussian noise. Assuming a noise-limited mmWave system, the signal-to-noise ratio (SNR) at user $k$ is given by
\begin{equation}
\gamma_k = \frac{P_{\text{tx}} |h_k|^2}{N_0}
\end{equation}
The achievable data rate for user $k$ is computed as
\begin{equation}
R_k = \log_2 \left(1 + \gamma_k \right)
\end{equation}


\section{Proposed DRL-Based BS-RIS Placement}
In this section, we formulate the joint Base Station (BS) and Reconfigurable Intelligent Surface (RIS) placement problem as a Markov Decision Process (MDP). We then detail the proposed Deep Deterministic Policy Gradient (DDPG) algorithm, which is uniquely suited for this continuous optimization task, followed by the dynamic mobility model used to evaluate the network's robustness.

\subsection{Markov Decision Process (MDP) Formulation}
Unlike traditional grid-search or discrete reinforcement learning methods (e.g., DQN) that suffer from the curse of dimensionality, our approach operates entirely in a continuous space. The environment models an all-green campus with $M$ buildings and $K$ users. The MDP is defined by the tuple $(\mathcal{S}, \mathcal{A}, \mathcal{R})$ as follows:

\subsubsection{State Space (Observations)}
The state vector $s_t \in \mathcal{S}$ is explicitly constructed to provide the agent with both geometric awareness and real-time network performance feedback. At any time step $t$, the state is a concatenated vector comprising:
\begin{itemize}
\item \textbf{Normalized Spatial Coordinates:} The 2D coordinates of all BSs and RISs across the $M$ buildings, contributing $4M$ values.
\item \textbf{Aggregated Network Statistics:} The minimum, maximum, and mean user data rates, alongside Jain's fairness index, contributing $4$ values.
\item \textbf{Individual User Rates:} The normalized data rates for all $K$ users to allow the agent to identify specific coverage holes.
\end{itemize}
Consequently, the total dimension of the state space is $4M + 4 + K$. Normalizing these inputs ensures stable gradient updates during the neural network training phase.

\subsubsection{Action Space}
The agent must determine the optimal locations for each BS and its corresponding RIS. Thus, the action vector $a_t \in \mathcal{A}$ is a continuous vector defined as:
\begin{equation*}
\begin{aligned}
a_t = [&x_{\text{BS},1},\, y_{\text{BS},1},\, x_{\text{RIS},1},\, y_{\text{RIS},1}, \\
      &\dots,\,
      x_{\text{BS},M},\, y_{\text{BS},M},\,
      x_{\text{RIS},M},\, y_{\text{RIS},M}]
\end{aligned}
\label{eq:action_space}
\end{equation*}
Because the raw output of the DDPG actor network spans an unbounded continuous space, we apply a geometry-aware mapping function to the environment step. First, the BS coordinates are mapped (snapped) to the valid dilated boundary of their respective buildings. Second, the RIS coordinates are snapped to the building's convex hull. Finally, a clamping constraint forces a minimum physical separation distance $d_{\min}$ between a BS and its corresponding RIS to prevent near-field physical collisions and hardware interference.

\subsubsection{Reward Function}
The optimization objective is to maximize network fairness and protect the most vulnerable links (max-min fairness). Therefore, the base reward is the worst-case (minimum) user Signal-to-Noise Ratio (SNR) in decibels. To strictly enforce the minimum separation constraint without breaking the continuous nature of the optimization, we apply a linear penalty for any distance violations. The reward function is given by:
$$r_t = \min_{k \in \mathcal{K}} \text{SNR}_k^{(\text{dB})} - \sum_{m=1}^M \rho \cdot \max(0, d_{\min} - d_{m})$$
where $d_{m}$ is the actual Euclidean distance between the $m$-th BS and RIS, and $\rho$ is a scaling factor. If all BS-RIS pairs satisfy the separation constraint, the penalty becomes zero, and the agent solely maximizes the minimum SNR.

\begin{algorithm}[t]
\caption{BS-RIS Location Optimization using Continuous DDPG}
\begin{algorithmic}[1]
\State \textbf{Input:} Users $\{(x_k,y_k)\}_{k=1}^K$, $M$ buildings masks, $d_{\min}$
\State Initialize actor $\pi_{\theta}$, critic $Q_{\phi}$, target networks $\pi_{\bar{\theta}}, Q_{\bar{\phi}}$, replay buffer $\mathcal{D}$
\For{each episode}
    \State Reset environment and obtain state $s_t$
    \For{each time step $t=1, \dots, T_{\max}$}
        \State Action selection with noise: $a_t = \pi_{\theta}(s_t) + \mathcal{N}$
        \For{each building $m \in \{1, \dots, M\}$}
            \State Extract $(x_{\text{BS},m}, y_{\text{BS},m}, x_{\text{RIS},m}, y_{\text{RIS},m})$ from $a_t$
            \State Snap BS coordinates to valid internal building mask
            \State Snap RIS coordinates to valid external convex hull mask
            \State Clamp RIS location to ensure $d_m \geq d_{\min}$
        \EndFor
        \State Compute physical channel SNRs and separation penalties
        \State Calculate reward $r_t$ and observe next state $s_{t+1}$
        \State Store $(s_t,a_t,r_t,s_{t+1})$ in $\mathcal{D}$
        
        \State Sample mini-batch of size $B$ from $\mathcal{D}$
        \State Update critic $Q_{\phi}$ using Bellman target $y$
        \State Update actor $\pi_{\theta}$ using policy gradient: $\nabla_{\theta} J \approx \nabla_{a} Q_{\phi}(s, a) \nabla_{\theta} \pi_{\theta}(s)$
        \State Soft update targets: $\bar{\theta} \leftarrow \tau \theta + (1-\tau)\bar{\theta}$, $\bar{\phi} \leftarrow \tau \phi + (1-\tau)\bar{\phi}$
    \EndFor
\EndFor
\State \textbf{Output:} Optimal continuous coordinates for all BSs and RISs
\end{algorithmic}
\end{algorithm}

\subsection{Handover Management}
While the previous section addresses optimal static BS/RIS placement, real users are mobile, and serving link quality changes continuously as they move between coverage areas, requiring an efficient handover mechanism to ensure seamless connectivity throughout the campus.

\subsubsection{User Mobility Model}
Users move throughout the campus following a waypoint mobility model, while the infrastructure remains fixed. Each user is randomly assigned an initial position and destination within the simulation grid, moving toward it at a constant walking speed $v$ and updating position at each timestep according to
\begin{equation}
\mathbf{p}_i(t+1) = \mathbf{p}_i(t) + v \, \Delta t \cdot \frac{\mathbf{d}_i(t) - \mathbf{p}_i(t)}{\lVert \mathbf{d}_i(t) - \mathbf{p}_i(t) \rVert}
\label{eq:waypoint_update}
\end{equation}
Upon arrival, the user pauses for a randomized dwell time $\tau_i \sim \mathcal{U}(\tau_{\min}, \tau_{\max})$ before a new destination is selected.
\begin{equation}
\tau_i \sim \mathcal{U}(\tau_{\min}, \tau_{\max})
\label{eq:dwell_time}
\end{equation}

To capture short-lived link degradation distinct from the slowly varying shadow fading in Section~\ref{sec:System}, transient blockage events representing momentary line-of-sight obstruction by pedestrians or vehicles are additionally modeled. Based on measured human-body blockage attenuations of 10--25~dB reported in outdoor and indoor line-of-sight measurement campaigns \cite{MacCartney2017,Huang2020}, blockage is modeled as a per-user, per-timestep Bernoulli event with probability $p = 0.03$ and attenuation $L_{\text{block}} = 15$~dB, applied to the target link's SNR during handover execution as
\begin{equation}
\gamma_{i,\text{target}}^{\text{dB}}(t) = \gamma_{i,\text{target}}^{\text{dB,nom}}(t) - B_i(t) \cdot L_{\text{block}}
\label{eq:blockage_snr}
\end{equation}

\begin{figure*}
    \centering
    \includegraphics[width=\linewidth]{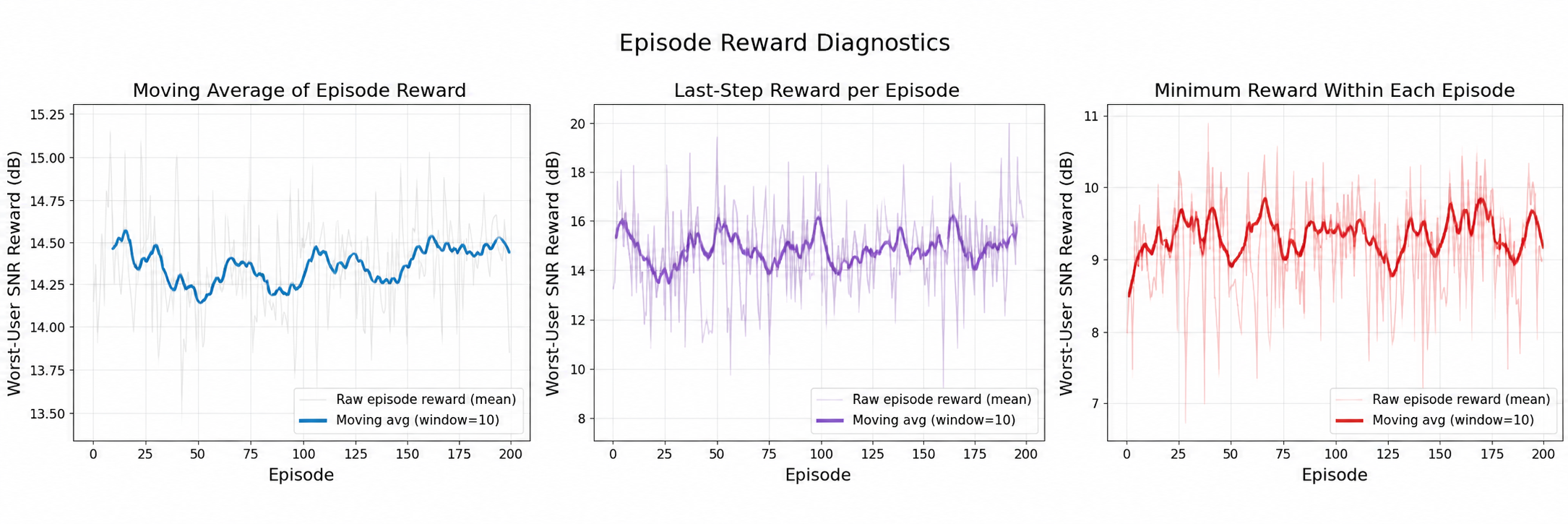}
    \caption{DDPG training diagnostics showing the moving-average reward, final reward per episode, and minimum reward observed within each episode.}
    \label{reward}
\end{figure*}

\begin{figure}[!t]
    \centering
    \includegraphics[width=\linewidth]{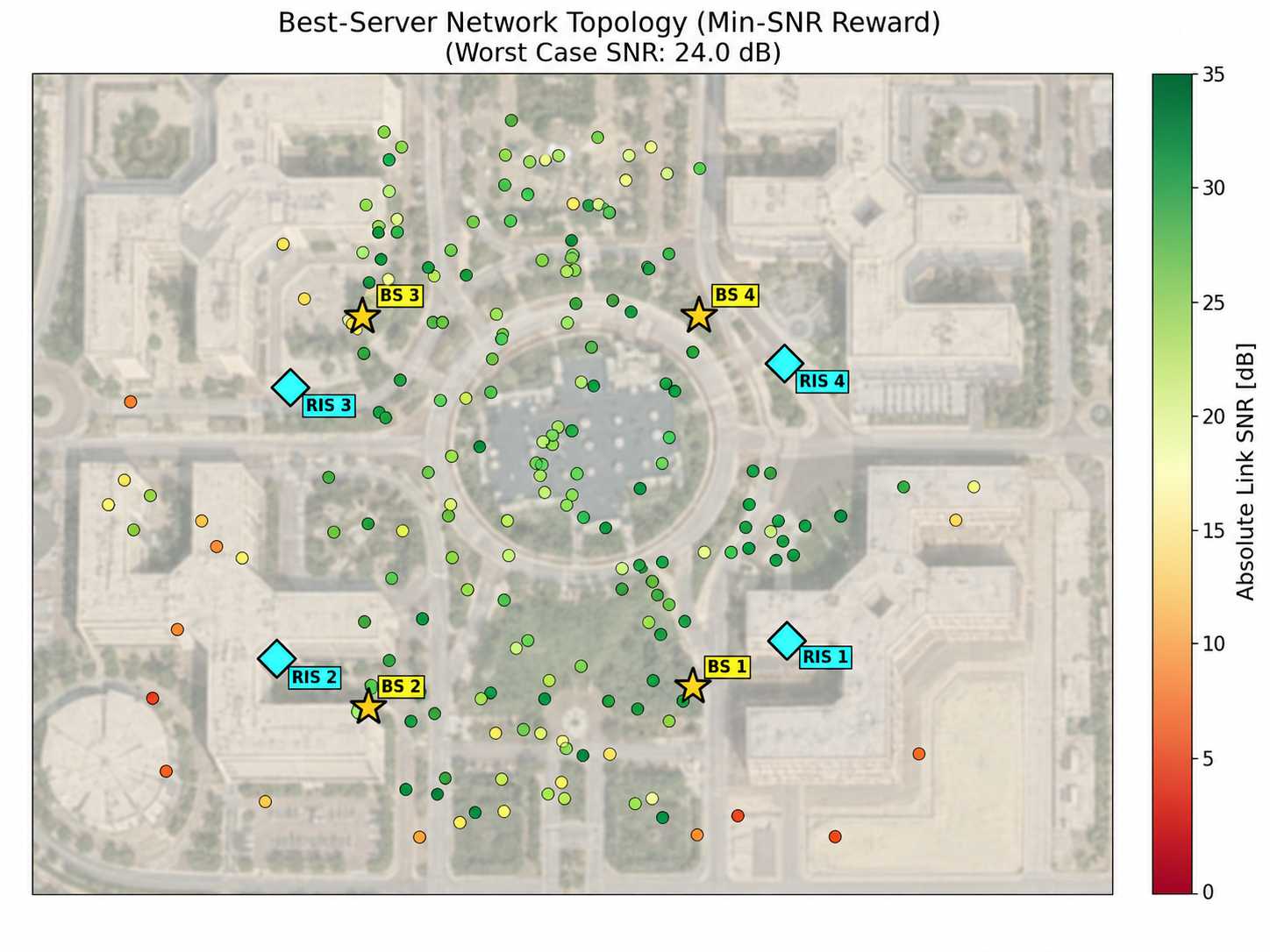}
    \caption{Users SNR Scale and Coverage Distribution}
    \label{pmddpg}
\end{figure}

\begin{figure}[!t]
    \centering
    \includegraphics[width=\linewidth]{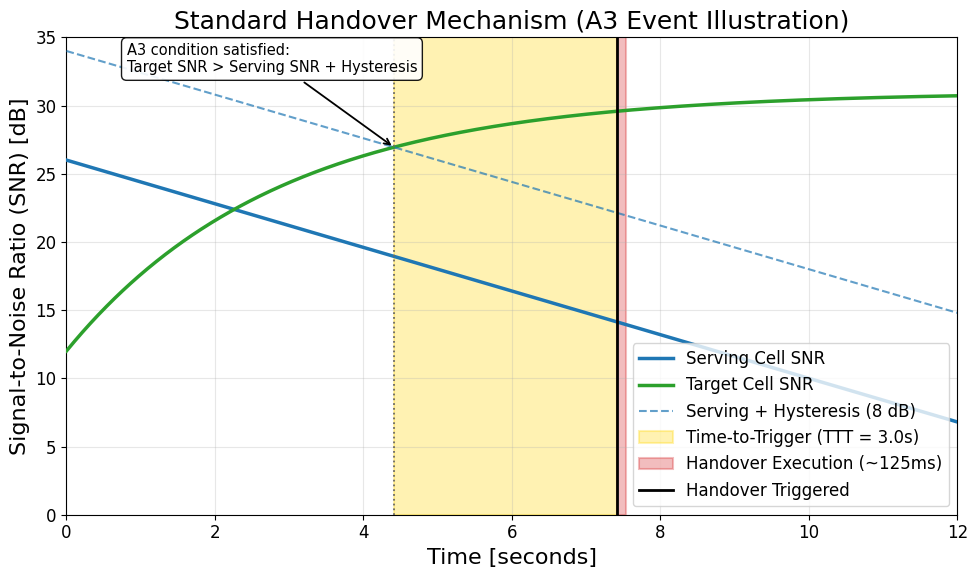}
    \caption{Illustration of the standard 3GPP A3-event handover mechanism, including hysteresis, Time-to-Trigger (TTT), handover execution, and triggering instant.}
    \label{fig:placeholder}
\end{figure}

\subsubsection{Handover Decision Mechanism}
Following the 3GPP A3 event definition, a target cell is considered only once its SNR exceeds the serving cell's SNR by a hysteresis margin $H$ for a fixed Time-to-Trigger (TTT) interval:
\begin{equation}
\gamma_{\mathrm{target}} > \gamma_{\mathrm{serving}} + H
\label{eq:handover_condition}
\end{equation}
If the condition is violated at any point during this interval, the timer resets and the user remains connected to the serving base station, reducing ping-pong effects and avoiding interrupted service. As the investigated system consists of closely-spaced buildings sharing a common pedestrian plaza, representative of ultra-dense small-cell HetNets, the adopted hysteresis and TTT values are elevated relative to macro-cell networks, consistent with prior findings that small-cell networks require case-based handover parameter selection \cite{Adaptive}.

\subsubsection{Handover Procedure}
Once \eqref{eq:handover_condition} is satisfied over the required TTT interval, a hard, break-before-make handover is triggered, with randomized execution latency of 0.5--2.0~s. The target link's SNR is monitored throughout this window rather than only at completion, since radio link failure results from the worst instantaneous channel condition during execution, not the final received signal quality \cite{MacCartney2017}. A handover succeeds only if the target link's SNR remains above a minimum threshold throughout execution; otherwise, the user enters a fixed-duration outage, during which no further handover may be triggered.

\subsubsection{Performance Metrics}
Handover performance is evaluated using the total handover attempts, successful handovers, success rate, average latency, and outage fraction -- the proportion of simulated time spent in a post-failure disconnected state, complementing the success rate.

\section{Numerical Results}

The proposed approach was evaluated through software-based simulations considering
$K=180$ users distributed across $M=4$ buildings/cells, with a minimum BS--RIS
separation distance of $d_{\min}=35$~m. The wireless channel model assumes a
transmit power of $P_{\mathrm{tx}}=30$~dBm (1~W), a noise power of
$N_0=-95$~dBm, a path-loss exponent of $\alpha=3.5$, a channel bandwidth of
20~MHz, and a carrier frequency of 10~GHz.

The DDPG agent was trained for 20{,}000 timesteps, with a maximum of 200 steps
per episode. The actor and critic learning rates were set to
$1\times10^{-4}$ and $1\times10^{-3}$, respectively, while the soft-update
factor and discount factor were set to $\tau=0.005$ and $\gamma=0.99$.
A replay buffer of 100{,}000 transitions and a batch size of 128 were used.
Both the actor and critic networks employed an MLP architecture of
$[128,128]$.

For the mobility evaluation, 180 users followed waypoint-based pedestrian
mobility for 0.5~min, corresponding to 600 simulation steps. The pedestrian
velocity was set to $1.4$~m/s. The handover procedure employed an A3-event
trigger with a hysteresis margin of $H_{\mathrm{dB}}=8$~dB and a
Time-to-Trigger (TTT) of 10 simulation steps. The TTT was selected to
represent the time required by a pedestrian moving at $1.4$~m/s to cross a
representative inter-cell boundary. Following TTT expiration, the target link
was continuously monitored until the handover was completed.

The performance was evaluated using the minimum, mean, and maximum user SNRs,
achievable data rates, Jain's fairness index, coverage percentage, handover
success rate, handover rate, and post-failure outage duration. Coverage was
evaluated for quality-of-service (QoS) thresholds of 50 and 100~Mbps.

Fig.~2 presents the training diagnostics of the DDPG agent. The left-hand
plot shows the moving average of the episode reward, which remains stable at
approximately 14.4~dB, indicating convergence without significant training
instability. The middle plot presents the reward obtained at the final step
of each episode. The observed fluctuations are attributed to variations in
user locations and stochastic wireless channel conditions; nevertheless, the
rewards remain within a consistent range throughout training.

The right-hand plot presents the minimum reward observed during each episode,
representing the worst-case network performance encountered during the
exploration of different BS and RIS placements. Although occasional drops are
observed under challenging user distributions and channel realizations, the
minimum reward generally remains above 7~dB. This behavior indicates that the
learned policy maintains the performance of the weakest users while preserving
a stable worst-case SNR objective.

The spatial distribution of the users and the resulting data rates obtained
with the learned DDPG policy are illustrated in Fig.~\ref{pmddpg}. The
continuous action-space formulation enables the agent to continuously adjust
the BS coordinates within the building boundaries while constraining the RIS
locations to the external convex hulls of the corresponding structures.
Consequently, the resulting deployment satisfies the considered physical
constraints without requiring grid-based coordinate quantization.

The optimized topology produces a favorable spatial distribution of the user
data rates. In particular, the learned policy places the BSs and RISs such
that the minimum BS--RIS separation constraint is satisfied while the RISs
are positioned to improve connectivity in regions with weaker coverage. This
spatial adaptation contributes to the resulting SNR, throughput, and fairness
performance.

Fig.~\ref{fig:placeholder} illustrates the A3-event triggering mechanism and
the corresponding handover execution process. The results show that a total
of five handover attempts occurred during the mobility simulation, and all
five attempts were successfully completed. This corresponds to a 100%
handover success rate and a handover rate of 0.06 handovers/user/min.

Furthermore, the post-failure outage duration was 0.00
user-steps, indicating that no connectivity interruption was observed during
the evaluated mobility scenario.

The optimized topology achieved an average SNR of 33.4~dB, while the
worst-user and best-user SNRs were 24~dB and 54.84~dB, respectively. The
resulting Jain's fairness index was 0.963, indicating a relatively uniform
distribution of the achieved user performance.

In addition, 100
QoS thresholds. These results demonstrate that the learned BS/RIS deployment
provides high network coverage while maintaining favorable SNR and fairness
performance under the considered static and pedestrian-mobility scenarios.

\section{Conclusion}
This paper proposes a DDPG framework for continuous 2D BS and STAR-RIS placement in a 6G campus network, enforcing a Max-Min Fairness objective to eliminate coverage dead zones and maximize worst-case SNR. Simulation results demonstrate that the learned deployment achieves strong worst-case SNR performance, high fairness, complete QoS coverage, and reliable mobility support through a standard A3-event handover mechanism with a 100\% success rate under the evaluated scenario.

\bibliographystyle{IEEEtran}
\bibliography{Ref}

\end{document}